\documentstyle[12pt]{article}
\newcommand{\p}[1]{\partial #1}
\newcommand{\dx}[1]{\int d^4 x #1}

\newcommand{\dk}[1]{\int {d^4 k \over (2\pi)^4 }#1}

\begin{document}
\begin{flushright}
NSC/USTC-30/96\\[8mm]
\end{flushright}
\begin{center}
{\Large\bf Electromagnetic Masses of the Massive \\
Yang-Mills Particles $K^*(892)$}\\[5mm]
Mu-Lin Yan and Dao-Neng Gao\\
{\small
Chinese Center for Advanced Science and Technology(World Lab),\\
P.O.Box 8730, Beijing, 100080, P.R.China\\
Center for Fundamental Physics,
University of Science and Technology of China\\
Hefei, Anhui 230026, P.R.China\footnote{Mailing Address}}\\
\end{center}
\begin{abstract}
\noindent
Electromagnetic mass difference between neutral $K^*$ and charged $K^*$
has been calculated in the $U(3)_L\times U(3)_R$ chiral fields theory
of mesons. It has been revealed that the non-abelian gauge structure
of the massive Yang-Mills lagrangian obeyed by $K^*$ 
plus VMD (vector meson dominace) causes the EM-mass
of neutral one larger than charged one. Experiment supports this
effect.
\end{abstract}

Since the Yang-Mills (YM) fields theory was discovered$^{[1]}$, several
important concepts and pictures based on the non-abelian gauge symmetry
structure of YM theory have been deeply rooted in the
particle physics. In this present letter we try to illustrate another
unusual effect of the massive Yang-Mills (MYM) theory obeyed by the mesons 
with spin one$^{[2]}$. Specifically, it will be revealed
below that the non-abelian gauge structure of MYM with
gauging Wess-Zumino-Wittern anomaly (GWZW)$^{[3]}$ (or
Bardeen anomaly $^{[4]}$) and VMD (vector meson dominace)$^{[5]}$ makes
the electromagnetic mass (i.e., electromagnetic self-energy) of
neutral $K^*(892) $(i.e.,$K^{*0}$ or $\bar{K}^{*0}$) larger than one
of charged $K^*$ (i.e., $K^{*\pm}$), namely
\begin{equation}
    m(K^{*0})_{EM}>m(K^{*+})_{EM}
\end{equation}
where the subscript $EM$ denotes electromgnetic mass. This is really unusual
because it is contrary to the common knowledges of the hadron's 
EM-masses, such as $m({\rm neutron})_{EM}<m({\rm proton})_{EM}^{[6]},
 m(\pi^0)_{EM}<m(\pi^+)_{EM}^{[7]}$, and $m(K^0)_{EM}<m(K^+)_{EM}^{[8]}$.
The practical calculations (see below)
will indicate that the contributions coming from MYM
lagrangian of $K^*$ plus VMD are significantly larger than one from
GWZW plus VMD. So the claim (1) is due to the non-abelian guage structure of
YM theory basically.
Using the result of $m(K^{*0})_{EM}-m(K^{*+})_{EM}$ calculated in this letter
and a known estimation of $(m(K^{*0})-m(K^{*+}))_{non-EM}\;^{[9]}$,
the total mass difference between $K^{*0}$ and $K^{*+}$ is predicted.
The result is in good agreement with the data, so eq.(1) has been supported experimentally.
 
The physics for low-lying mesons ($\pi, K, \eta, \eta ',\rho, \omega
, a_1, f, K^*, \phi, K_1, f_s$) is very rich. The dynamics guiding them
can be constructed in terms of the principles of non-perturbative QCD
theory. A reliable and satisfactory theory for these mesons is required
to meet all constraints coming from both the experimental data and
the theoretical considerations, and to pay the parameters as few as 
possible. Recently such a theory called $U(3)_L\times U(3)_R $ 
chiral fields theory of pseudoscalar, vector, and axial-vector mesons
$^{[10]}$ has been proposed and studied for various mesonic processes.
This theory provides an unified description of meson physics at low
energies. The basic inputs for it are the cutoff $\Lambda$ (or $g$ in
[10]) and $m$ (related to quark condensate). The phenomenologies of the
theory are quite encouraging. The error bars for most of the processes
are less than 20 percent, which are consistent with the approximation of large 
$N_c$ expansion. Theoretically, this theory has many attractive features.
We would like to emphasize some of them here as follows:

1, The VMD is a natural
consequence of this theory instead of an input. According to ref.[10],
VMD reads
\begin{eqnarray}
{\cal L}_{\rho \gamma}=-{e \over f_\rho} \partial_\mu
    \rho_\nu^0 (\partial^\mu A^\nu -\partial^\nu A^\mu) \\
{\cal L}_{\omega \gamma}=-{e \over f_\omega} \partial_\mu
    \omega_\nu (\partial^\mu A^\nu -\partial^\nu A^\mu) \\
{\cal L}_{\phi \gamma}=-{e \over f_\phi} \partial_\mu
    \phi_\nu (\partial^\mu A^\nu -\partial^\nu A^\mu) 
\end{eqnarray}
The direct couplings of photon to other mesons are obtained
through the substitutions
\begin{equation}
\rho_\mu^0 \longrightarrow 
               {e \over f_\rho} A_\mu, \;\;\;\;
\omega_\mu \longrightarrow {e \over f_\omega} A_\mu, \;\;\;
\phi_\mu \longrightarrow {e \over f_\phi} A_\mu
\end{equation}
where
\begin{equation}
{1\over f_\rho}={1\over 2}g,\;\;\;
{1\over f_\omega}={1\over 6}g,\;\;\;
{1\over f_\phi}=-{1\over 3\sqrt{2}}g.
\end{equation}
We choose the parameter $g=0.38\pm 0.01$ in this letter, which corresponds to
taking the experimental value $m_a=1.23\pm 0.04GeV^{[11]}$
as an input [10].

2, This theory starts with a chiral lagrangian of quantum quark fields within 
mesonic background fields, and
the chiral dynamics of the mesons
comes from the path integration over quark fields.
Thus a cutoff $\Lambda$ due to quark loop calculations has to be
introduced$^{[10]}$ into this truncated effective fields theory as follows,
\begin{equation}
g^2={8\over 3}{ N_{c} \over (4\pi)^{D/2}} {D \over 4}
({\mu^2 \over m^2})^{\epsilon /2} \Gamma (2-{D\over 2})
\end{equation}
($D=4-\epsilon$) or in terms of cut-off $\Lambda $ 
\begin{equation}
g^2={8\over 3}{ N_{c} \over (4\pi)^{2}}
\{ log(1+{\Lambda^2 \over m^2})
+{1 \over 1+{\Lambda^2 \over m^2}}-1\}. 
\end{equation}
Here the $g$ (or $\Lambda$) emerges as an intrinsic parameter and
serves to describe the ultraviolet logarithm divergence in the theory.
Therefore, it is legitimate to use the $g$ (or $\Lambda$) to factorize
the logarithm divergence in the loop calculations of this truncated
fields theory. 

3, Due to the universality of couplings in this theory, all the couplings
in the lagrangian are fixed by $g$ and $m$. Therefore, no new parameter
needs to be introduced into the GWZW anomaly part of this theory.

Thus, the electromagnetic interactions of mesons have been constructed
by VMD naturally and the divergences of photon-meson loops can be
factorized consistently, then the calculations of meson's EM-masses
become practicable and reliable. We will present systematical studies
to all low-lying meson's EM-masses in detail elsewhere. 
In this letter we focus on the EM-masses of $K^{*}$.

Generally, in order to get the virtual photon contributions to the masses 
of mesons, we use ${\cal L}_i (\Phi ,\gamma,...)|_{\Phi=\pi, K, K^*,...}$ 
to calculate the following S-matrix
\begin{equation}
S_{\Phi}=\langle \Phi |T{\rm exp}[i\int dx^4 {\cal L}_i
(\Phi,\gamma,...)]-1|\Phi\rangle |_{\Phi=\pi, K, K^*,...}.
\end{equation}
On the other hand $S_\Phi$ can also be expressed in terms of
the effective lagrangian of $\Phi$ as
$$
S_\Phi=\langle \Phi |i\int d^4 x {\cal L}_{{\rm eff}} (\Phi) | \Phi
\rangle.
$$
Noting ${\cal L}={1\over 2}\p_\mu\Phi\p^\mu \Phi-{1\over 2}
m_{\Phi}^2 \Phi^2$, then the electromagnetic interaction
correction
to the mass of $\Phi$ reads
\begin{equation}
\delta m_\Phi^2 ={2iS_\Phi \over \langle \Phi |\Phi^2 |
\Phi \rangle }.
\end{equation}
where $\langle \Phi |\Phi^2 |\Phi \rangle =
\langle \Phi |\dx \Phi^2(x) |\Phi \rangle $. 

We adopt dimensional regularization to do loop-calculations,
and take the most generic linear gauge condition
for electromagnetic fields to all diagram calculations in
this letter. Namely, the $A_\mu$-propagator with an arbitrary
gauge constant $a$ is taken to be
\begin{eqnarray}
{\Delta_{F}^{(\gamma)}} _{\mu\nu}(x-y)&=&\dk
     {{\Delta_{F}} ^{(\gamma)}} _{\mu\nu}
     (k)e^{-ik(x-y)}, \nonumber \\
{\Delta_F^{(\gamma)}} _{\mu\nu}(k)&=&
 {-i \over k^2} [g_{\mu\nu}-(1-a){k_\mu k_\nu \over k^2}].
\end{eqnarray}

The calculations in this letter are of $O(\alpha_{em})$ and
one-loop. The interactions between $K^*$ and other mesons consist three
parts: gauging non-linear $\sigma$ model, MYM lagrangian and 
GWZW lagrangian $^{[10]}$. 
There are two sorts of vertices which contribute to $m^2(K^{*0})_{EM}-
m^2(K^{*+})_{EM}$: three points vertices and four points vertices. Due to
VMD, the former has to be the coupling of $K^*$ to a neutral vector meson plus
other field, and the latter is the interaction of two $K^*$ with two neutral
vector-mesons(Here, the neutral vector mesons are  $\rho^0$,
$\omega$ and $\phi$). It is easy to be sure that all contributions to $m^2(K^{*0})_{EM}
-m^2(K^{*+})_{EM}$ come from MYM- and GWZW- parts of $K^*-$lagrangian, and
no contributions are from gauging $\sigma-$model. In the following, we calculate
them separately.

From[10], MYM lagrangian related to $K^*$ reads
\begin{equation}
{\cal L}_{MYM}=-{g^2 \over 8}Tr(\p_\mu v_\nu-\p_\nu v_\mu-i[v_\mu ,v_\nu]
-i[a_\mu, a_\nu])^2 +{\rm mass\; terms\; of}\; v,
\end{equation}
where $v_\mu=\tau_i \rho^i_\mu+\lambda_aK^{*a}_\mu
+({2\over 3}+{1\over \sqrt{3}}\lambda_8)\omega_\mu
+({1\over 3}-{1\over \sqrt{3}}\lambda_8)\phi_\mu $, and $a_\mu$ is
axial vector field ( notations of [10] are used ). The three-point MYM vertex
contrbuting to $K^*$'s EM masses
and four-point one read from eq.(12)
\begin{eqnarray}
{\cal L}_{MYM}^{(3)}(K^{*\pm})=-{i\over g}(\rho^0_{\mu\nu}+\omega_{\mu\nu}-
\sqrt{2}\phi_{\mu\nu})K^{*+\mu}K^{*-\nu} \nonumber \\
+{i\over g}(\rho^0_\mu+\omega_\mu-\sqrt{2}\phi_\mu)(K^{*+\mu\nu}K^{*-}_\nu
   -K^{*-\mu\nu}K^{*+}_\nu), \\
{\cal L}_{MYM}^{(3)}(K^{* 0})=-{i\over g}(-\rho^0_{\mu\nu}+\omega_{\mu\nu}-
\sqrt{2}\phi_{\mu\nu})K^{*0\mu}\bar{K}^{*0\nu}\nonumber \\
+{i\over g}(-\rho^0_\mu+\omega_\mu-\sqrt{2}\phi_\mu)
(K^{*0\mu\nu}\bar{K}^{*0}_\nu
   -\bar{K}^{*0\mu\nu}K^{*0}_\nu), \\
{\cal L}_{MYM}^{(4)}(K^{*\pm})=-{1\over g^2}(\omega^{\mu}-\sqrt{2}\phi^{\mu})
 \{ 2\rho^0_\mu K^{*+\nu}K^{*-}_\nu \nonumber \\
     -\rho^{0\nu}(K^{*+}_\mu K^{*-}_{\nu}+K^{*+}_{\nu}K^{*-}_\mu)\},\\
{\cal L}_{MYM}^{(4)}(K^{*0})={1\over g^2}(\omega^{\mu}-\sqrt{2}\phi^{\mu})
 \{ 2\rho^0_\mu K^{*0\nu}\bar{K}^{*0}_\nu \nonumber \\
 -\rho^{0\nu}(K^{*0}_\mu \bar{K}^{*0}_{\nu}+K^{*0}_{\nu}\bar{K}^{*0}_\mu)\},
\end{eqnarray}
where $v_{\mu\nu}=\p_\mu v_\nu-\p_\nu v_\mu ,\;\;(v=\rho,\;\omega,\;
\phi,\;K^*)$. The corresponding direct couplings of $K^*$ to photon,
${\cal L}^{(3)}_{MYM}(K^{*\pm},\gamma),
 {\cal L}^{(4)}_{MYM}(K^{*\pm},\gamma), 
 {\cal L}_{MYM}^{(4)}(K^* v\gamma)|_{v=\rho, \omega, \phi} $
 and ${\cal L}^{(3)}_{MYM}(K^{*0},\gamma)
 ={\cal L}^{(4)}_{MYM}(K^{*0},\gamma)=0$,
can be obtained through the VMD substitutions eq.(5) respectively.
Thus the S-matrix is
\begin{equation}
S=\langle K^*|T\{{\rm exp}\;i\int d^4x[{\cal L}_{MYM}(K^*)
+{\cal L}_{MYM}(K^*\gamma)+{\cal L}_{\rho \gamma}+{\cal L}_{\omega \gamma}
+{\cal L}_{\phi \gamma}]-1\}|K^*\rangle
\end{equation}
which can be calculated in the standard way.Then the EM-mass is given by formula of (10).

The Feynman diagrams corresponding to ${\cal L}^{(3)}_{MYM}$ and
${\cal L}^{(4)}_{MYM}$ are shown in Fig.(1) and Fig.(2) respectively.

Firstly, let's check the gauge-independence of the S-matrix. The gauge-dependent
part of the S-matrix of Fig.(1) has been derived from eqs.(11), (17) and
${\cal L}^{(3)}_{MYM}(K^{*\pm}\gamma)$, which is
\begin{equation}
S^{(3)G}_{MYM}={3a \over 4}e^2 \langle K^{*+} K^{*-}\rangle
\int {d^4 k \over (2\pi)^4}{1\over k^2},
\end{equation}
where $\langle K^{*+} K^{*-}\rangle =\int d^4 x 
\langle K^{*+}|{1\over 2}(K_1^{*2}+K_2^{*2})| K^{*-}\rangle $.
Similarly, the gauge-dependent part of Fig.(2) (corresponding to
${\cal L}^{(4)}_{MYM}(K^{*\pm}\gamma)$) is
\begin{equation}
S^{(4)G}_{MYM}={-3a \over 4}e^2 \langle K^{*+} K^{*-}\rangle
\int {d^4 k \over (2\pi)^4}{1\over k^2},
\end{equation}
From eqs.(18),(19), we have $S^{(3)G}_{MYM}+S^{(4)G}_{MYM}=0$.
Therefore the EM-mass calculations of MYM part is gauge-independent.
Furthermore, when one takes the dimensional regularization,
$S^{(3)G}_{MYM}=S^{(4)G}_{MYM}=0 $, then $S^{(3)}_{MYM}$ and $S^{(4)}_{MYM}$
become gauge-independent individually. To the GWZW anomaly part,
since there is an antisymmetric tensor $\epsilon^{\mu\nu\lambda\rho}$
in the lagrangian (see below), the gauge-independency is obvious. So
we conclude that our EM-mass calculations in this letter is 
gauge-independent.

From Fig.(1), the EM-mass square difference
corresponding to the contributions of three-point MYM vertices
theory can be calculated directly. The result is
\begin{eqnarray}
(m_{EM}^2(K^{*0})&-&m_{EM}^2(K^{*\pm}))_{MYM}^{(3)}= 
{-ie^2\over 3}\int {d^4k\over (2\pi)^4}m_\rho^2 \nonumber \\
&\times& {1\over -k^2(-k^2+2p\cdot k)
(-k^2+m_\rho^2)} \nonumber \\
&\times&({m_\omega^2 \over -k^2+m_\omega^2}+{2m_\phi^2 \over -k^2+m_\phi^2})
\nonumber\\
&\times &\{ 3k^2+4m_{K^*}^2-{k^4\over m_{K^*}^2}
-4{(k\cdot p)^2 \over k^2}+2k\cdot p \nonumber \\
&&+{\langle (k\cdot K^{*+})(k\cdot K^{*-}) \rangle \over
\langle K^{*+}K^{*-}\rangle} (9+{k^2 \over m_{K^*}^2}-{2k\cdot p\over k^2})\},
\end{eqnarray}
where $p$ is 4-momentum of $K^*$, i.e., $p^2=m_{K^*}^2$. It is standard
to carry the Feynman integration in eq.(20) through to the end. The logarithmic
divergences in it are factorized by eq.(7). Then we have
\begin{equation}
(m_{EM}^2(K^{*0})-m_{EM}^2(K^{*\pm}))_{MYM}^{(3)}
=\Delta_\omega +\Delta_\phi
\end{equation}
where
\begin{eqnarray}
\Delta_\omega &=&{e^2m_\rho^2 m_\omega^2 \over 12\pi^2}\{ {45m_{K^*}^2 \over
16m_\rho^2m_\omega^2}-{49m_{K^*}^4(m_\rho^2+m_\omega^2) \over 180m_\rho^4
m_\omega^4} \nonumber \\
&+&\int_0^1 dx{1\over m_\omega^2-m_\rho^2}[{m_{K^*}^2\over m_\omega^2}
(Y_\omega ({5\over 2}x-{7\over 2})-1-2x^2+x^3)logY_\omega \nonumber \\
 & &-{m_{K^*}^2\over m_\rho^2}(Y_\rho({5\over 2}x-{7\over 2})-1-2x^2+x^3)
logY_\rho \nonumber \\
&&+{m_{K^*}^4\over m_\omega^4}({1\over 8}Y_\omega^2+{1+x^2 \over 2}Y_\omega
+x^2)log Y_\omega \nonumber \\
&&-{m_{K^*}^4\over m_\rho^4}({1\over 8}Y_\rho^2+{1+x^2 \over 2}Y_\rho
+x^2)log Y_\rho ]\} \nonumber \\
&-&{e^2m_\rho^2m_\omega^2 \over m_{K^*}^2}
({1 \over 32}g^2+{1 \over 64\pi^2}log{f_\pi^2 \over 6(g^2 m_\rho^2-f_\pi^2)}
\nonumber \\
&&+{1\over 48\pi^2}-{1\over 64\pi^2}{m_\omega^2 \over m_\omega^2-m_\rho^2}
log{m_\omega^2 \over m_\rho^2}) \nonumber \\
&+&{e^2\over 64\pi^2}{m_\rho^2 m_\omega^2 \over m_\omega^2-m_\rho^2}
log{m_\omega^2 \over m_\rho^2}, \\
\Delta_\phi &=& 2\Delta_\omega (m_\omega^2 \rightarrow m_\phi^2),
\end{eqnarray}
where
$$ Y_i=x^2+{m_i^2 \over m_{K^*}^2}(1-x),\;\;i=\rho,\;\omega,\;\phi. $$
Substituting the experimental values of $f_\pi(=0.186GeV), \;
m_\rho,\; m_\omega, \; m_\phi $ and $m_{K^*}$ into eqs.(21) (22) and (23), we get
the numerical result
\begin{equation}
(m_{EM}^2(K^{*0})-m_{EM}^2(K^{*\pm}))_{MYM}^{(3)}
=1.57\pm 0.02\times 10^{-3} GeV^2.
\end{equation}

Similarly, the four-point MYM-vertex contributions to EM-mass difference
of $K^*$ (Fig.(2))
can be calculated by using ${\cal L}_{MYM}^{(4)}(K^*),
{\cal L}_{MYM}^{(4)}(K^* \gamma)$,
${\cal L}_{MYM}^{(4)}(K^* v\gamma)|_{v=\rho, \omega, \phi},
{\cal L}_{\rho \gamma},
{\cal L}_{\omega \gamma}$ and ${\cal L}_{\phi \gamma} $.
It is straightforward to get the result as follows
\begin{eqnarray}
\lefteqn{(m_{EM}^2(K^{*0})-m_{EM}^2(K^{*\pm}))_{MYM}^{(4)}}\nonumber \\
&=&{3e^2 \over 64\pi^2} \{{m_\rho^2 m_\omega^2 \over m_\omega^2-m_\rho^2}
log {m_\omega^2 \over m_\rho^2}
+2{m_\rho^2 m_\phi^2 \over m_\phi^2-m_\rho^2}
log {m_\phi^2 \over m_\rho^2}\}.
\end{eqnarray}
Numerically, we have
\begin{equation}
(m_{EM}^2(K^{*0})-m_{EM}^2(K^{*\pm}))_{MYM}^{(4)} 
=9.38\times 10^{-4}GeV^2.
\end{equation}

In the GWZW anomaly part of the $U(3)_L\times U(3)_R $ chiral theory
of mesons there are vertices of $K^*-(\rho,\omega,\phi)-K$. So this
part makes contributions to the EM-masses of $K^*$, which should be
taken into account. The anomaly lagrangians for $K^{*\pm}$ and $K^{*0}$
read respectively
\begin{eqnarray}
{\cal L}_{GWZW}(K^{*\pm})
&=&-{N_c \over 2\pi^2 g^2 f_K}\epsilon^{\mu\nu\alpha\beta}
(K_\mu^{*+} \p_\beta K^-+K^{*-}_\mu \p_\beta K^+) \nonumber \\
&&\times ({1\over2}\p_\nu \rho^0_\alpha+{1\over2}\p_\nu \omega_\alpha
+{\sqrt{2}\over2}\p_\nu \phi_\alpha ), \\
{\cal L}_{GWZW}(K^{*0})
&=&-{N_c \over 2\pi^2 g^2 f_K}\epsilon^{\mu\nu\alpha\beta}
(K_\mu^{*0} \p_\beta \bar{K}^0+\bar{K}^{*0}_\mu \p_\beta K^0) \nonumber \\
&&\times(-{1\over2}\p_\nu \rho^0_\alpha+{1\over2}\p_\nu \omega_\alpha
+{\sqrt{2}\over2}\p_\nu \phi_\alpha ),
\end{eqnarray}
where $f_K$ is determined by the following equation,
$$ {f_K^2 \over 1-{f_K^2 \over g^2m_{K^*}^2}}=
{f_\pi^2 \over 1-{f_\pi^2 \over g^2m_{\rho}^2}}
$$
The corresponding direct couplings of $K^*$ to photon,
 ${\cal L}_{GWZW}(K^{*\pm},\gamma)$ and ${\cal L}_{GWZW}(K^{*0},\gamma)$,
can be obtained through the VMD substitutions of eq.(5). 
Then, the GWZW's contribution to
the EM-mass difference of $K^*$ can be computed directly. The result is
\begin{eqnarray}
(m_{EM}^2(K^{*0})-m_{EM}^2(K^{*\pm}))_{GWZW}
&=&{-9e^2\over 2\pi^4 g^2 f_K^2}\int_0^1 dx_1\int_0^{x_1}dx_2\int_0^{x_2}
dx_3 \nonumber \\
& \times& m_{K^*}^2{m_\rho^2\over 48\pi^2}({m_\omega^2\over M_a^2}
-{2m_\phi^2 \over M_b^2}),
\end{eqnarray}
where
\begin{eqnarray*}
M_a^2&=&m_\omega^2x_3+m_\rho^2(x_2-x_3)+m_K^2(1-x_1)-m_{K^*}^2x_1(1-x_1),\\
M_b^2&=&m_\phi^2x_3+m_\rho^2(x_2-x_3)+m_K^2(1-x_1)-m_{K^*}^2x_1(1-x_1).
\end{eqnarray*}
Numerically, we have
\begin{equation}
(m_{EM}^2(K^{*0})-m_{EM}^2(K^{*\pm}))_{GWZW} 
=6.95\pm 0.33\times 10^{-4}GeV^2.
\end{equation}
The full EM-mass difference between $K^{*0}$ and $K^{*\pm}$ is the sum of
eqs.(24),(25) and (30), which reads
\begin{equation}
(m({K^{*0}})-m({K^{*\pm}}))_{EM}=1.79\pm 0.03MeV.
\end{equation}
In above computations, there are no any adjustable parameters.
In other words, when one takes the experimental value of 
$m_\rho, m_a, m_{K^*}, m_K, m_\omega, m_\phi $ and $f_\pi $ as
inputs, then $(m({K^{*0}})-m({K^{*\pm}}))_{EM}$ is fixed.
The fact that this value is positive is just the claim of eq.(1).
We can see also that the contribution of MYM part is significantly 
larger than one of GWZW part,
\begin{equation}
(m({K^{*0}})-m({K^{*\pm}}))_{EM}^{MYM}:(m({K^{*0}})-m({K^{*\pm}}))_{EM}^{GWZW}
=3.75:1.
\end{equation}
Therefore, under VMD, this EM-mass difference is dominated by the
MYM lagrangian of $K^*$.

The total mass difference between $K^{*0}$ and $K^{*\pm}$ should be
expressed as
\begin{eqnarray}
\lefteqn{(m(K^{*0})-m(K^{*\pm}))_{{\rm total}}}  \nonumber \\
&=&(m(K^{*0})-m(K^{*\pm}))_{non-EM}+(m(K^{*0})-m(K^{*\pm}))_{EM},
\end{eqnarray}
where $(m(K^{*0})-m(K^{*\pm}))_{non-EM}$ is non-electromagnetic 
contrbution to the mass difference. As the quark masses enter the effective
lagrangian, this non-electromagnetic contribution can be estimated. 
In ref.[9], such an estimation has been done and the authors have
made the best recommendation to the value of this quantity as follows
\begin{equation}
(m(K^{*0})-m(K^{*\pm}))_{non-EM}=4.47MeV.
\end{equation}
Then the prediction is
\begin{equation}
(m(K^{*0})-m(K^{*\pm}))_{total}=6.26\pm0.03MeV .
\end{equation}
which is in good agreement with the experimental data $6.7\pm 1.2 MeV
\;^{[11]}$. This fact means that the experiment favors to support the
theoretical prediction of eq.(31) even though the experimental error
is still large so far.

To conclude: In terms of the $U(3)_L\times U(3)_R$ chiral fields theory
of mesons, all one-loop diagrams contributing to the electromagnetic
mass difference of $K^*$ have been calculated. The calculations are
gauge independent. The gauging non-$\sigma$ model part of the theory
has no contribution to EM-masses of $K^*$ while MYM- and GWZW-
parts make contributions. MYM- and GWZW-parts make the EM-mass
of $K^{*0}$ larger than one of $K^{*\pm}$ respectively, but MYM-part
is dominant. Then we conclude that in this reliable theory of mesons
the non-abelian gauge fields structure of Yang-Mills theory causes 
the EM-mass of neutral $K^*$ larger than one of charged $K^*$. This
effect of YM theory is remarkable. It is interesting that the experiment
supports it.

Similar studies can be done to other vector mesons and such an effect
will also emerge.  However,
the things will not be so clear as in the $K^*$ case
becouse the non-linear $\sigma$
model part of the meson's dynamics will be involved in the evaluations.
We will provide such an investigation elsewhere.

\begin{center} {\bf ACKNOWLEDGMENTS} \end{center}
We would like to thank Bing An Li for helpful discussions. This work
was supported in part by the National Science Funds of China through
Chen Ning Yang.


\vspace{10mm}
\newpage
\leftline{\bf Caption}
\begin{description}
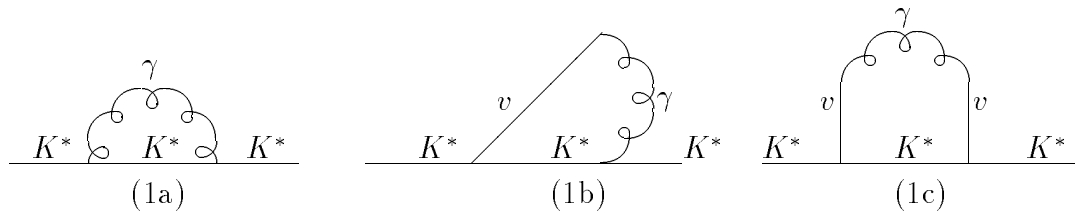
\item[Fig.1] $S_{MYM}^{(3)}$-Feynman diagrams corresponding to
      ${\cal L}_{MYM}^{(3)}(K^*)
({\rm eqs.(13)(14)}),\\
 {\cal L}_{MYM}^{(3)}(K^*\gamma)$,
and ${\cal L}_{\gamma\rho}$, ${\cal L}_{\gamma\omega}$,
${\cal L}_{\gamma\phi}$ (eqs.(2)-(4)),
the curly line is photon-line, $v$ denotes
neutral vector mesons $\rho^0$,$\omega$ and $\phi$.
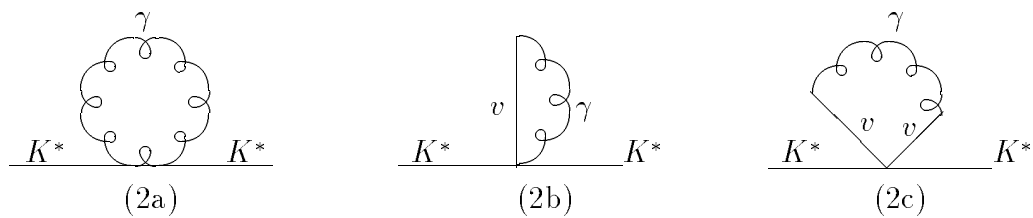
\item[Fig.2] $S_{MYM}^{(4)}$-Feynman diagrams corresponding to
${\cal L}_{MYM}^{(4)}(K^*)$
(eqs.(15)(16))
${\cal L}_{MYM}^{(4)}(K^* v\gamma)|_{v=\rho, \omega, \phi}$,
${\cal L}_{MYM}^{(4)}(K^*\gamma), $
and ${\cal L}_{\gamma\rho}$, ${\cal L}_{\gamma\omega}$,
${\cal L}_{\gamma\phi}$ (eqs.(2)-(4)),
the curly line is photon-line, $v$ denotes
neutral vector mesons $\rho^0$,$\omega$ and $\phi$.
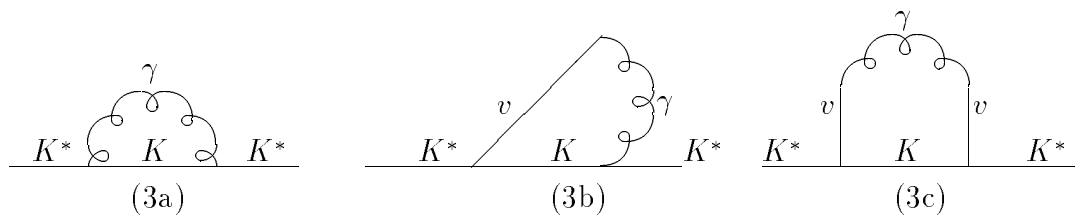
\item[Fig.3] $S_{GWZW}$-Feynman diagrams corresponding to
${\cal L}_{GWZW}(K^*)  \;
({\rm eqs.(27)(28)}),\\
 {\cal L}_{GWZW}(K^*\gamma)$,
and $ {\cal L}_{\gamma\rho}, {\cal L}_{\gamma\omega}, 
{\cal L}_{\gamma\phi}$ (eqs.(2)-(4)),
the curly line is photon-line, $v$ denotes
neutral vector mesons $\rho^0$,$\omega$ and $\phi$.
\end{description}


\begin{thebibliography}{40}
\bibitem{1} C.N.Yang and R.L.Mills, Phys.Rev.,{\bf 96},191(1954).
\bibitem{2} J.Schwinger, Phys.Lett., {\bf B24}, 473 (1967);
  J.Wess and B.Zumino, Phys.Rev.,{\bf 163}, 1727 (1967);
  S.Weinberg, ibid. {\bf 166}, 1568 (1968);
  B.W.Lee and H.T.Nieh, ibid. {\bf 166}, 1507 (1968).
\bibitem{3} J.Wess and B.Zumino, Phys.Lett., {\bf B37}, 95 (1971);
E.Wittern, Nucl.Phys., {\bf B223}, 422 (1983);
\"{o}. Kaymakcalan, S.Rajeev and J.Schechter, Phys. Rev., 
{\bf D30}, 594 (1984);
H.Gomm, \"{o}. Kaymakcalan and J.Schechter, Phys. Rev., 
{\bf D30}, 2345 (1984);
Chou Kuang-Chao, Guo Han-Ying, Wu Ke and Song Xing-Chang, Phys.Lett., {\bf B134}, 67 (1984).
\bibitem{4} W.A.Bardeen, Phys.Rev., {\bf 184}, 1848 (1969). 
\bibitem{5} J.J.Sakurai, $Currents\;and\;Mesons$, (University of Chicago
Press, Chicago, 1969).
\bibitem{6} B.A.Li, M.L.Yan and K.F.Liu, Phys.Lett., {\bf B177}, 409 (1986).
\bibitem{7} T.Das, G.Guralnik, V.Mathur, F.Low and J.Young
 Phys. Rev. Lett.,{\bf 18}, 759 (1967) ;
 J.Gasser and H.Leutwyler, Phys. Rep., {\bf 87},
 77 (1982);
 R.D.Peccei and J.Sola, Nucl. Phys.,{\bf B281},1
(1987); G.Ecker, J.Gasser, A.Pich and E.de Rafael, ibid. {\bf B321}, 311
(1989); 
 J.Bijnens and E.d.Rafael, Phys. Lett.,
{\bf B273},483(1991);
A.Duncan, E.Eichten and H.Thacker, Phys.Rev.Lett., {\bf 76}, 3894 (1996).
\bibitem{8} R.Dashen, Phys. Rev.,{\bf 183}, 1245(1969); 
 J.F.Donoghue, B.R.Holstein and D. Wyler, Phys. Rev. Lett., {\bf 69}, 3444
 (1992); J.F.Donoghue, B.R.Holstein and D.Wyler, Phys. Rev.,{\bf D47}
, 2089 (1993);  
 R.Baur and R.Urech, Phys. Rev, {\bf D53}, 6552 (1996).
\bibitem{9} J.Schechter, A.Subbaraman and H.Weigel, Phys. Rev. {\bf D48},
339 (1993).
\bibitem{10}B.A.Li, Phys. Rev.,{\bf D52}, 5165(1995).
\bibitem{11}Particle Data Group, Phys. Rev.{\bf D50}
No.3 (1994).
\end{thebibliography}
\end{document}